\documentclass[twocolumn]{aastex701}

\begin{document}
\shorttitle{Globular cluster IMF from Streams}
\shortauthors{Ye \& Carlberg}

\title{Inferring Globular Cluster Initial Mass Function from Stellar Streams}

\correspondingauthor{Claire S.\ Ye}
\author[0000-0001-9582-881X]{Claire S.\ Ye}
\affiliation{Canadian Institute for Theoretical Astrophysics, University of Toronto, 60 St. George Street, Toronto, Ontario M5S 3H8, Canada}
\email[show]{claireshiye@cita.utoronto.ca}

\author[0000-0002-7667-0081]{Raymond G.\ Carlberg}
\affiliation{Department of Astronomy \& Astrophysics, University of Toronto, Toronto, ON M5S 3H4, Canada}
\email{raymond.carlberg@utoronto.ca}

\begin{abstract}
The Gaia mission has provided precise astrometry and spectrophotometry for billions of stars in the Milky Way, enabling the identification and kinematic characterization of stellar streams. These streams, remnants of disrupted globular clusters and dwarf galaxies, have revealed the structure of the Milky Way's dark matter halo. We show that stellar streams also encode information about the initial mass function of globular clusters. We combine cold dark matter simulations that model the evolution and disruption of embedded globular clusters with observations of stellar streams and globular clusters to infer the initial cluster mass function. We find that initially more massive clusters produce more massive streams, but deposit a smaller fraction of their initial mass into those streams. Using stream mass and angular momentum measurements, we recover a declining, power-law-like initial mass function with a slope $\alpha = 1.4\pm0.05$ for streams $\gtrsim 1000\,M_{\odot}$ (with a maximum value of $\sim 1.6$ if we strictly use lower limits of stream mass measurements). This work establishes stellar streams as a novel probe of the early mass distribution of globular clusters.
\end{abstract}

\section{Introduction}\label{sec:intro}
The discovery and characterization of stellar streams in the Milky Way halo have provided unique insights into the hierarchical structure formation and dark matter content of our Galaxy. These streams are the tidal debris of globular clusters and dwarf galaxies that have dissolved within the Galactic potential \citep[e.g.,][]{Lynderbell_Lyndenbell_1995}, thereby also preserving a record of star cluster and galaxy formation in the early Universe.

In particular, thin and dynamically cold stellar streams are often identified as the tidal remnants of globular clusters, most notably Palomer~5 \citep[e.g.,][]{Odenkirchen+2001,Rockosi+2002}. The vast majority of known streams have a narrow metallicity and age distribution similar to those of individual halo globular clusters \citep[e.g.,][]{Bonaca_PW_2025}. However, it is important to note that most stream-finding algorithms are designed to target these features, potentially introducing selection biases into current stream catalogs \citep{Malhan2019,Ibata24}.

It is well understood that the present-day population of globular clusters is reduced in both mass and number from the population at early times as a consequence of stellar evolution, internal dynamical evolution, and tidal forces from the host galaxies \citep[e.g.,][]{Krumholz+2019,Kruijssen_2026}. Models assuming an initial cluster mass function (ICMF) similar to the low-redshift analogs of globular clusters, in which $dN/dM \propto M^{-2}$ from young massive star clusters \citep[e.g.,][and references therein]{PZ10,Krumholz+2019}, are able to successfully reproduce the currently observed log-normal globular cluster distribution in the Milky Way \citep[e.g.][]{Gnedin+2014,Reina-Campos+2022}. However, cluster formation histories and their initial properties remain poorly constrained at high redshifts. Recent observations by the James Webb Space Telescope (JWST) have provided the first glimpses of proto-globular clusters at redshift $z \gtrsim 2$ \citep[e.g.,][]{Mowla+2022,Vanzella+2022,Vanzella+2023,Adamo+2024,Rivera-Thorsen+2024,Senchyna+2024,Whitaker+2025}.

The combined population of stellar streams and old globular clusters offers an alternative approach for reconstructing the original cluster mass distribution. This reconstruction needs to account for several biases, including the completeness of observed stream and cluster catalogs, the mass fraction of progenitor clusters currently residing in the visible streams, and the population of clusters that have fully dissolved. While the census of Galactic globular clusters is fairly complete out to $\sim 100$~kpc \citep[e.g.,][]{Harris1996}, stellar streams are significantly harder to identify beyond $\sim 20$~kpc \citep[e.g.,][]{Pearson+2024} and are rapidly dispersed if they encounter the Galactic disk. On the other hand, clusters more massive than $\sim 10^5 M_\odot$ orbiting between 10-30~kpc can survive for a Hubble time, whereas their lower-mass counterparts typically dissolve entirely \citep{FallZhang}. Consequently, the observed stream mass represents only a fraction of the initial progenitor mass, and the current census of visible streams reflects only a fraction of the original cluster population.

In this work, we employ cosmological simulations to model the evolution of stellar streams and determine the necessary corrections to selection biases. We present the first analysis that uses the observed Galactic streams and globular clusters to predict the initial mass of their progenitor population. We describe the cosmological simulations in Section~\ref{sec:stream_sim} and the regression analysis on the simulations in Section~\ref{sec:estimate_mini}. We apply the regression fits to observational data to estimate the progenitor mass distribution in Section~\ref{sec:prediction}. Finally, we discuss the primary sources of uncertainties in Section~\ref{sec:uncer} and conclude in Section~\ref{sec:conclu}.

\section{Cosmological Simulations of Streams}\label{sec:stream_sim}

The cosmological simulations with embedded globular star particle clusters, initially developed in \citet{Carlberg18} and \citet{CK22}, are detailed in \citet{Carlberg24}. The analysis presented here uses the CDM simulations and includes an additional run with lower-mass star clusters. These lower-mass clusters have a smaller star particle softening length, but the simulations are otherwise identical. These simulations start from the same initial conditions using the \texttt{MUSIC} code \citep{MUSIC}. A low-resolution simulation was run in a 40/h~Mpc box to identify Milky-Way like galactic halos, which had at most a single massive galaxy companion within a 1/h Mpc volume and had no massive mergers below redshift one. A suitable region was then regenerated with \texttt{MUSIC} using dark matter particles of 10,322~$M_\odot$ and a total mass of $1.22 \times 10^{12}\,M_\odot$. The particles were evolved to a time of 1~Gyr when halos and subhalos are identified with the AHF code \citep{AHF1,AHF2}. Star clusters are inserted into the halos with masses greater than $3\times 10^8 M_\odot$ \citep{Benitez-Llambay20}, roughly the minimum mass for visible dwarf galaxies.

The globular clusters were split into two initial mass ranges, [$4\times10^4, 3\times 10^5 M_\odot$] and [$5\times10^3, 2\times 10^4 M_\odot$]. The upper mass range had $d\log{N}/d\log{M}=-1$ and the lower -0.5, which avoids over-populating the lower masses, but requires a statistical correction to the widely expected $d\log{N}/d\log{M}=-2$ relation \citep{PZ10}. The exact mechanisms of globular cluster formation are not yet fully understood. However, they are broadly theorized to form within self-gravitating gas in a rotating, disk-like distribution within dwarf-galaxy-like subhalos. This likely occurs approximately 1~Gyr after the Big Bang, when Milky Way-mass galactic halos were rare. Based on this premise, star particle clusters were initialized on exponentially distributed disk orbits at the local circular velocity, with an additional random velocity component of 5~${\rm km\,s^{-1}}$ applied in each direction, and in subhalos with masses above $2\times 10^8\,M_\odot$ with total cluster mass proportional to the dark matter mass \citep[e.g.,][]{Hudson+2014, Harris+2015}. The clusters were initiated as King models \citep{King1966} inside their tidal radii, resulting in clusters with half-mass radii of 5~pc $(M/3\times 10^5 M_\odot)^{1/3}$. The star particles have masses of $1\,M_\odot$ with a softening length of 1~pc for the lower mass range and 2~pc for the upper.

The streams are found using the great circle plane of their progenitor star clusters even if dissolved. The highest density ridgeline is found in galactic center spherical coordinates, in both velocity and position. The ridgeline is fitted with a fourth-order polynomial in stream latitude. The stream ends are set when the ridgeline sky density of a stream drops below 20~$M_\odot$ per square degree, which approximates the observational stream member finding limit. Streams are also required to have a minimum length of 10\degr. The stream masses are estimated by summing the star particles that are stream members.

Streams are produced over the lifetime of the progenitor cluster. Stars disperse into lower densities as gravitational encounters with dark matter subhalos increase the spread in angular momentum, leading to differential angular velocities and orbits. The cosmological simulation used here has vigorous merging up to about 7~Gyr, after which the inner halo potential evolves slowly while maintaining a population of orbiting subhalos. Streams are thus predominantly formed by stars stripped from their progenitors after 7~Gyr, provided the progenitors survived until that epoch. This process yields a set of streams with all-sky morphologies (e.g., the length distributions and orientations; \citealp{Carlberg18} and their Figure~16) consistent with the findings of the \texttt{STREAMFINDER} catalog in \citet[][]{Ibata24}. A detailed comparison between the simulated streams and the observations of \citet{Ibata24}, accounting for observational selection functions, is left for future work. Stream survival is lowest for low-mass progenitors and low-angular-momentum orbits, and vice versa (also see Section~\ref{sec:estimate_mini}).

\section{Estimating Initial Cluster Mass from Remnant}\label{sec:estimate_mini}

In this section, we fit the simulated masses and angular momenta of streams and remnant clusters using ordinary least-square (OLS) regression, providing the models for predicting initial cluster mass distribution from present-day stream and remnant cluster properties. All angular momentum values reported for both the simulations and observations refer to the total angular momentum.

\begin{figure}
\begin{center}
\includegraphics[width=0.9\columnwidth]{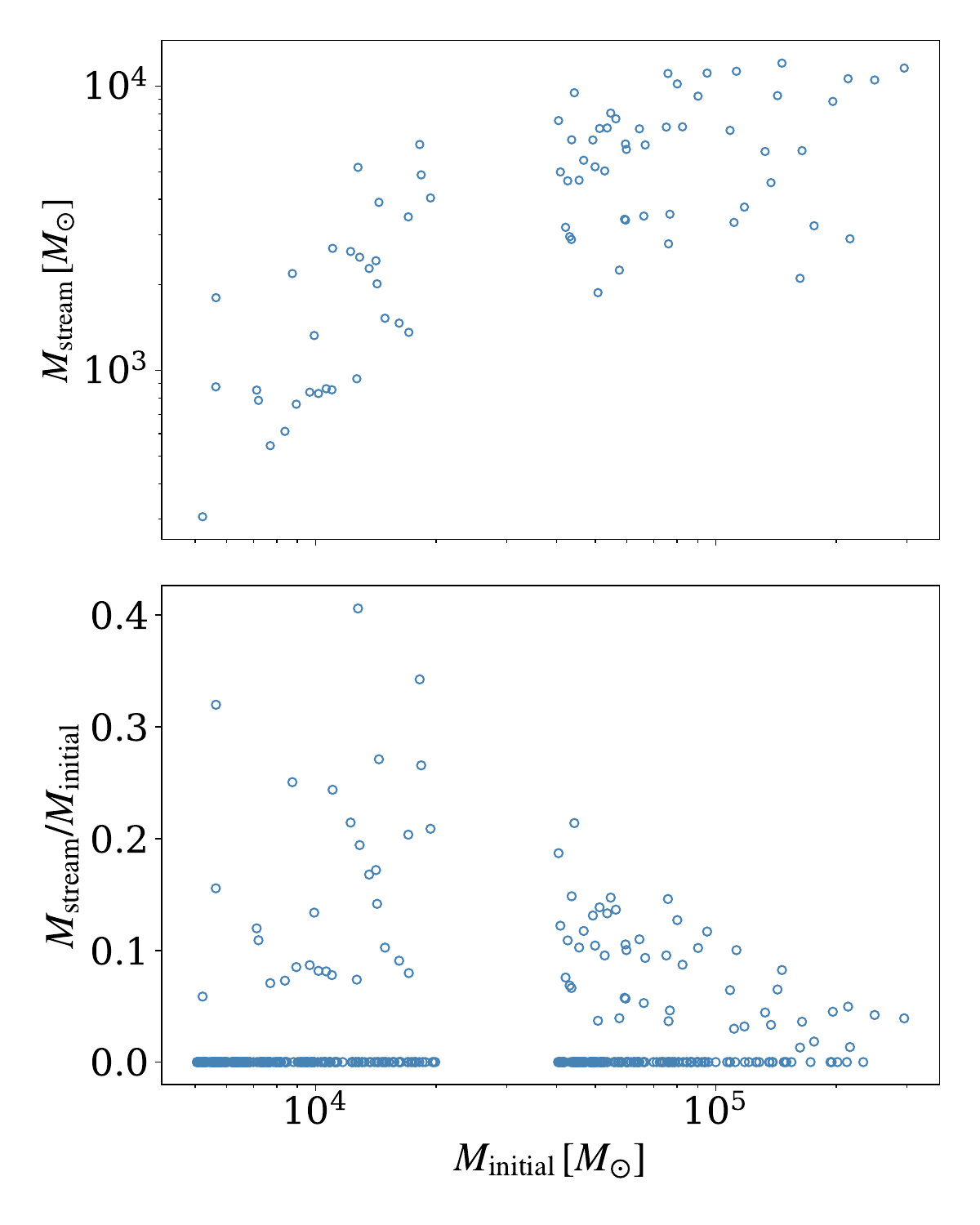}
\caption{Upper panel: Simulated stream mass as a function of initial cluster mass for streams visible at the present day. More massive clusters tend to leave behind more massive streams. Bottom panel: the fraction of mass remained in visible streams at the present day as a function of the initial cluster mass from simulations. A higher fraction of mass remains in streams from lower-mass star clusters.}\label{fig:mi_ms}
\end{center}
\end{figure}

Figure~\ref{fig:mi_ms} shows the visible stream mass as a function of initial cluster mass from the simulations in Section~\ref {sec:stream_sim}. The upper panel shows all visible stream masses, and the bottom panel shows the fraction of mass remaining in the visible streams, including systems with no visible stream at the present day, which primarily result from the dissolution of low-mass progenitor clusters during the Galaxy's active merging phase, widely dispersing their stars. As expected, there is a strong correlation between the initial cluster mass and the visible stream mass, suggesting that observed stream masses can be used to infer the ICMF of globular clusters. 

\begin{figure}
\begin{center}
\includegraphics[width=0.9\columnwidth]{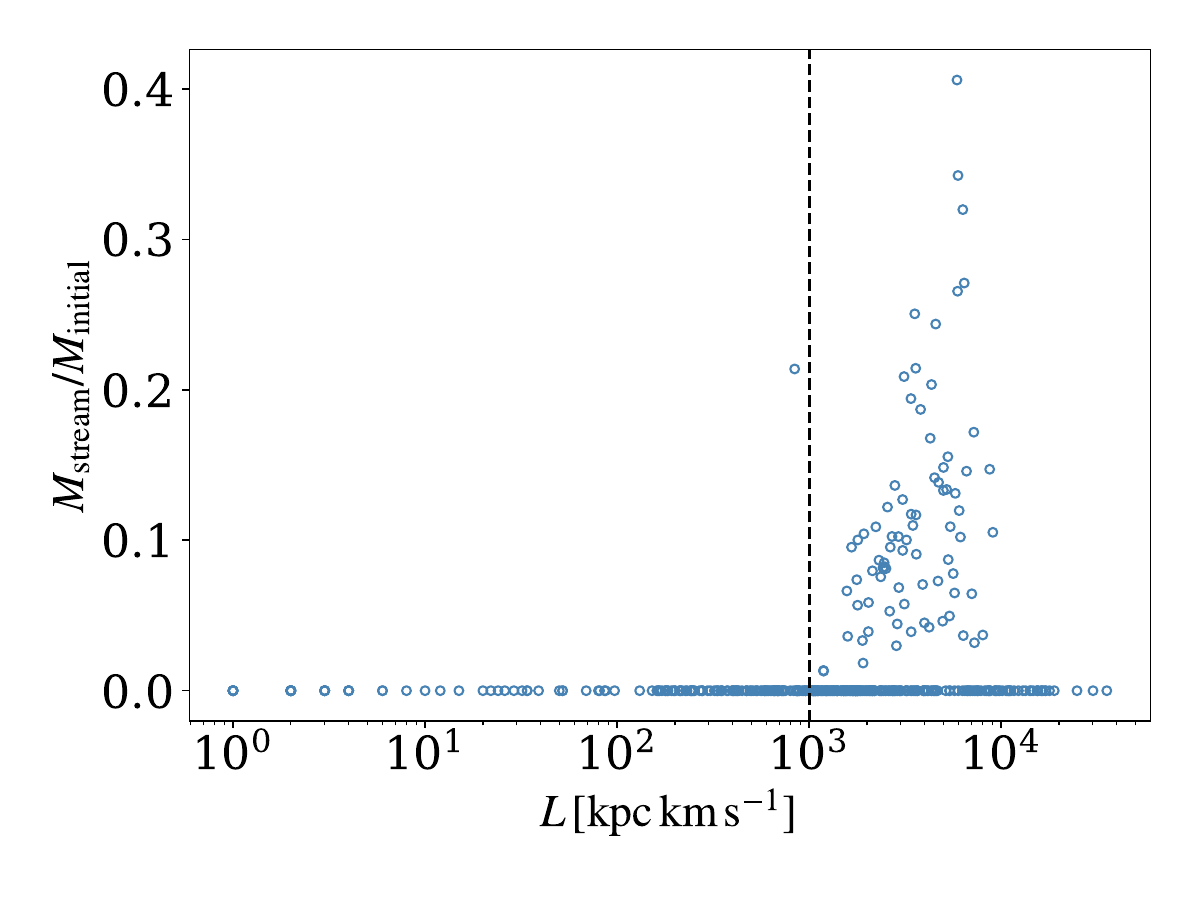}
\caption{Stream-to-initial mass ratio versus total angular momentum of the simulated streams at the present day. All simulated visible streams have total angular momentum $L \gtrsim 800 \text{ kpc km s}^{-1}$, a selection effect resulting from the preferential disruption of lower-angular-momentum subhalos by the tidal potential of the growing main halo and disk over the past $\sim8$~Gyr (disk potential from \citealt{galpy}). The vertical dashed line marks $1000\,{\rm kpc\,km\,s^{-1}}$.
}\label{fig:l_mr_stream}
\end{center}
\end{figure}

Many systems with low total angular momenta lack visible streams because their orbits subject the progenitor globular clusters to strong tidal forces from the Galactic disk, ultimately destroying the clusters and dispersing their integrated structures. We show the fraction of mass remained in streams versus their angular momentum in Figure~\ref{fig:l_mr_stream}, where almost all visible streams have $L>1000\,{\rm kpc\,km^{-1}}$. Note that the cosmological simulations discussed in Section~\ref{sec:stream_sim} search for visible streams for any angular momentum. Thus, angular momentum is a strong indicator of stream survival.

\begin{figure}
\begin{center}
\includegraphics[width=0.9\columnwidth]{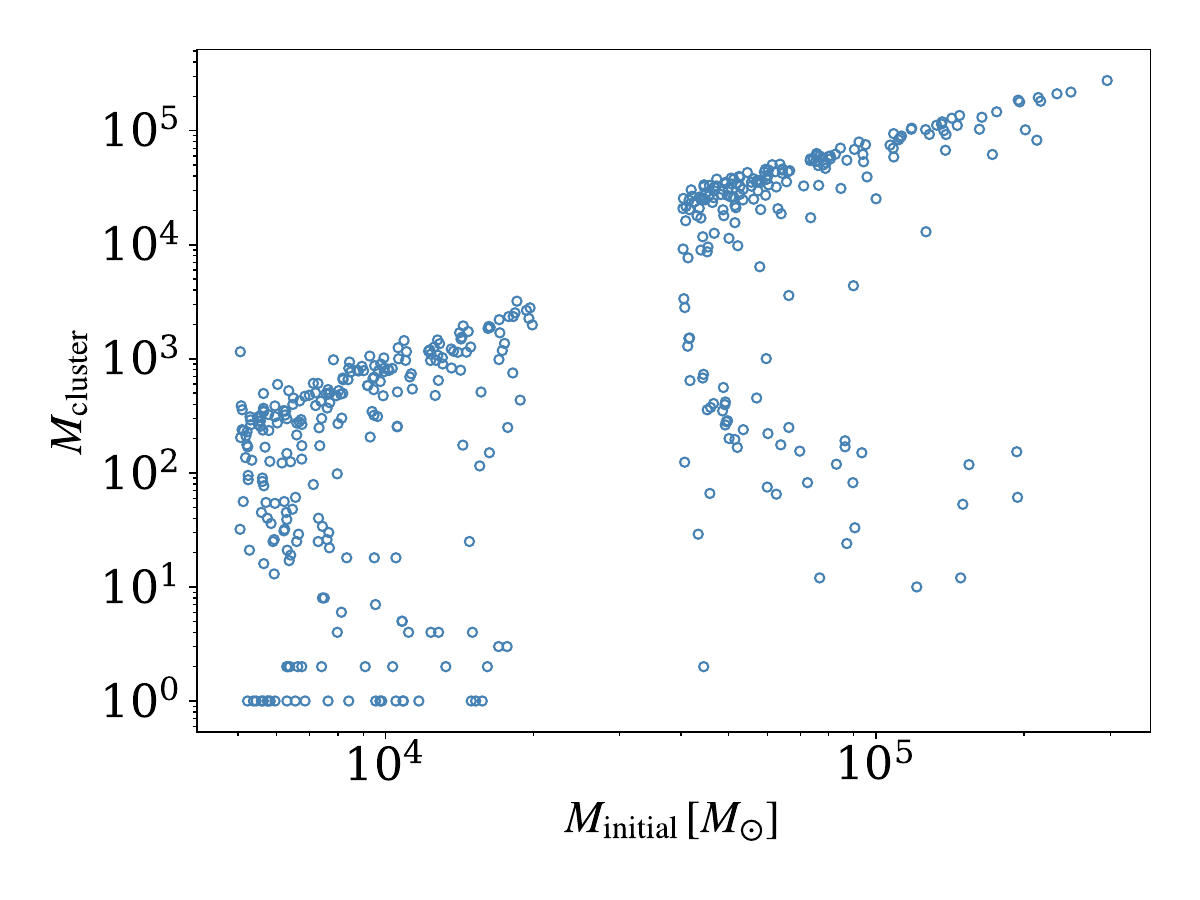}
\caption{Simulated remnant cluster mass as a function of initial cluster mass. Initially more massive clusters tend to leave behind more massive remnant clusters, but with a large scatter in the final cluster mass due to the difference in their angular momenta.}\label{fig:mi_mr}
\end{center}
\end{figure}

\begin{figure}
\begin{center}
\includegraphics[width=0.9\columnwidth]{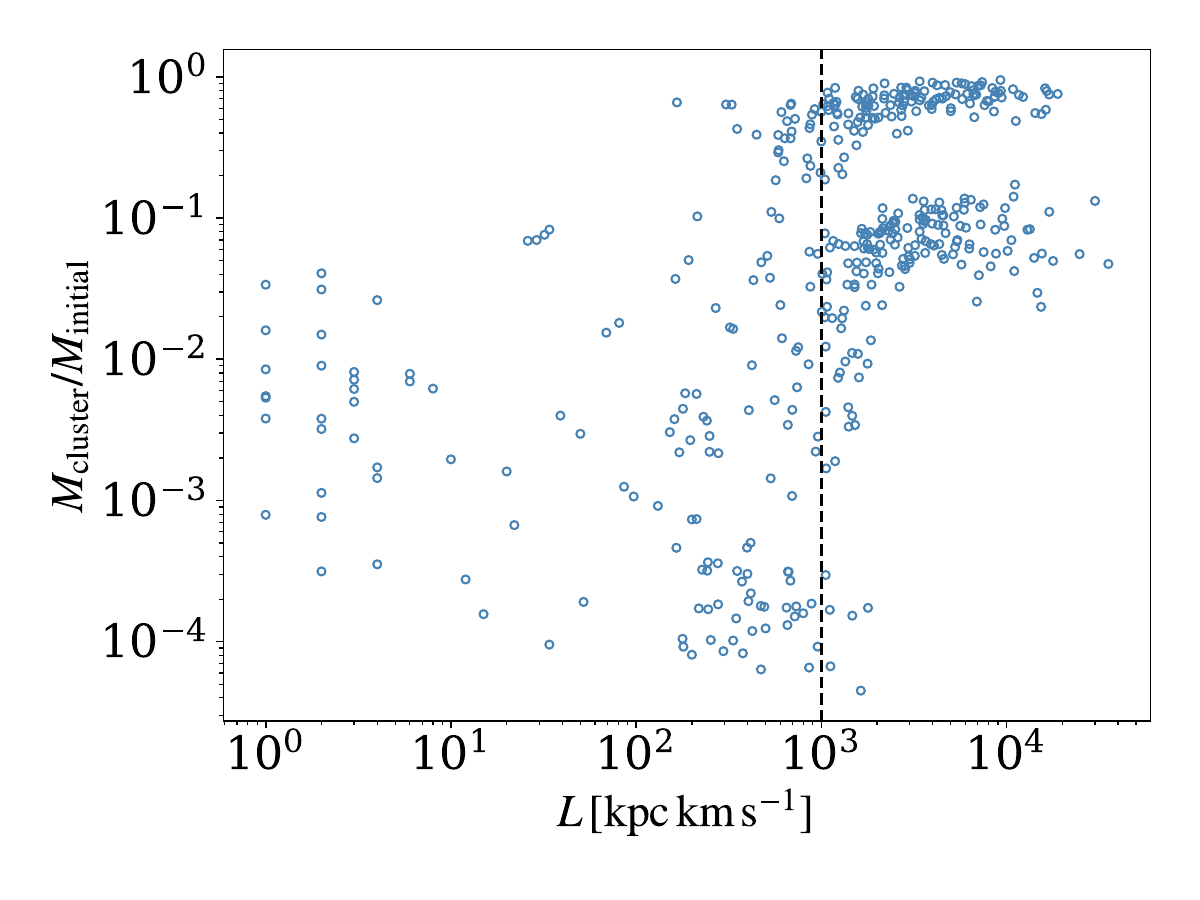}
\caption{Similar to Figure~\ref{fig:l_mr_stream}, but for remnant cluster-to-initial mass ratio versus total angular momentum at the present day. The initial clusters start to lose $\gtrsim 90\%$ mass when their angular momentum $L \lesssim 1000\,{\rm kpc\,km\,s^{-1}}$. The vertical dashed line marks $1000\,{\rm kpc\,km\,s^{-1}}$.}\label{fig:l_mr_cluster}
\end{center}
\end{figure}

We also show the final cluster mass versus the initial in Figure~\ref{fig:mi_mr} and the ratio of the remnant cluster mass to the initial cluster mass as a function of angular momentum in Figure~\ref{fig:l_mr_cluster}. The mass loss over the evolution histories of the clusters exhibits a similar dependence on angular momentum, with lower angular momentum and stronger tidal potential leading to more significant cluster dissolution. As with the streams, initially more massive clusters tend to retain more bound mass at the present day, with a large scatter reflecting their diverse angular momenta. Initially more massive clusters on high-angular-momentum orbits ($L \gtrsim 1000 \rm{\,kpc\,km\,s^{-1}}$) can retain over 10\% of their initial mass. In contrast, clusters across different mass ranges likely lose more than 90\% of their mass at lower angular momenta.

The stream and remnant cluster masses shown in Figures~\ref{fig:mi_ms} and \ref{fig:mi_mr} are independent of the choice of the simulations' ICMFs. While the simulations adopt a relatively shallow slope to efficiently sample the mass range, the resulting survival fractions can be convolved with any assumed ICMFs to recover the full initial population.

\subsection{Regression Analysis}\label{subsec:ols}
\begin{figure}
\begin{center}
\includegraphics[width=\columnwidth]{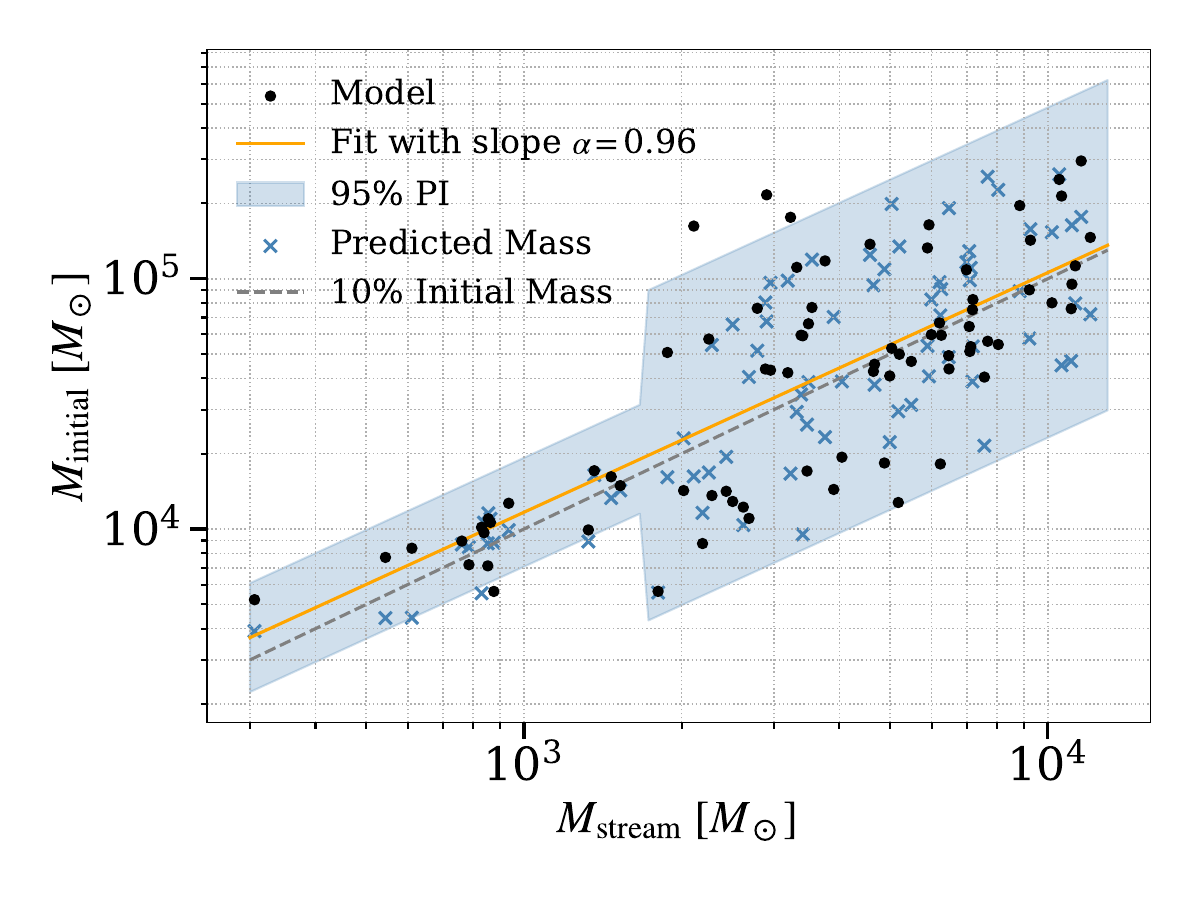}
\caption{Log-log fit to the simulated initial cluster mass versus the present-day stream mass using weighted OLS regression in two mass bins (boundary at $M_{\rm stream}=1700\,M_{\odot}$). Here we include only systems with $L>1000\,{\rm kpc\,km\,s^{-1}}$. The fit has a slope $\alpha=0.96$. The blue band shows the 95\% prediction interval based on empirical scatter, and the crosses show the possible predicted initial cluster mass based on the fit and the 95\% interval. The gray dashed line represents a reference case where $M_{\rm initial}= 10M_{\rm stream} $.}\label{fig:predictm_stream}
\end{center}
\end{figure}

Given the strong correlation of present-day stream and remnant cluster masses with initial mass and angular momentum, we perform a linear regression analysis to model this relationship. To derive the ICMF from visible stream and present-day cluster masses, our analysis proceeds in two stages: first accounting for the final-to-initial mass relation, and then for the overall survival fraction. We estimate the linear relation between ${\rm log}\,M_{\rm stream}$ and ${\rm log}\,M_{\rm initial}$ using weighted OLS, and show the fit and its 95\% prediction interval in Figure~\ref{fig:predictm_stream}. Here, we select only streams with angular momentum $L > 1000\,{\rm kpc\,km\,s^{-1}}$ from the cosmological simulations, since the ones with smaller angular momenta all have zero present-day visible mass except for one (Figure~\ref{fig:l_mr_stream}). The log-log fit has a slope of 0.96. The prediction intervals are weighted differently in two mass bins, with the boundary at $M_{\rm stream} = 1700~M_{\odot}$. This boundary is selected based on the change in the standard deviation of the simulation data in Figure~\ref{fig:predictm_stream}. The use of two distinct fits for the prediction intervals is a data-driven choice, though it coincides with the fact that the underlying simulations assume two different mass regimes (see also Figure~\ref{fig:mi_mr}). The lower-mass bin has a narrower prediction interval due to the small number of simulated visible streams in this regime.

\begin{figure}
\begin{center}
\includegraphics[width=\columnwidth]{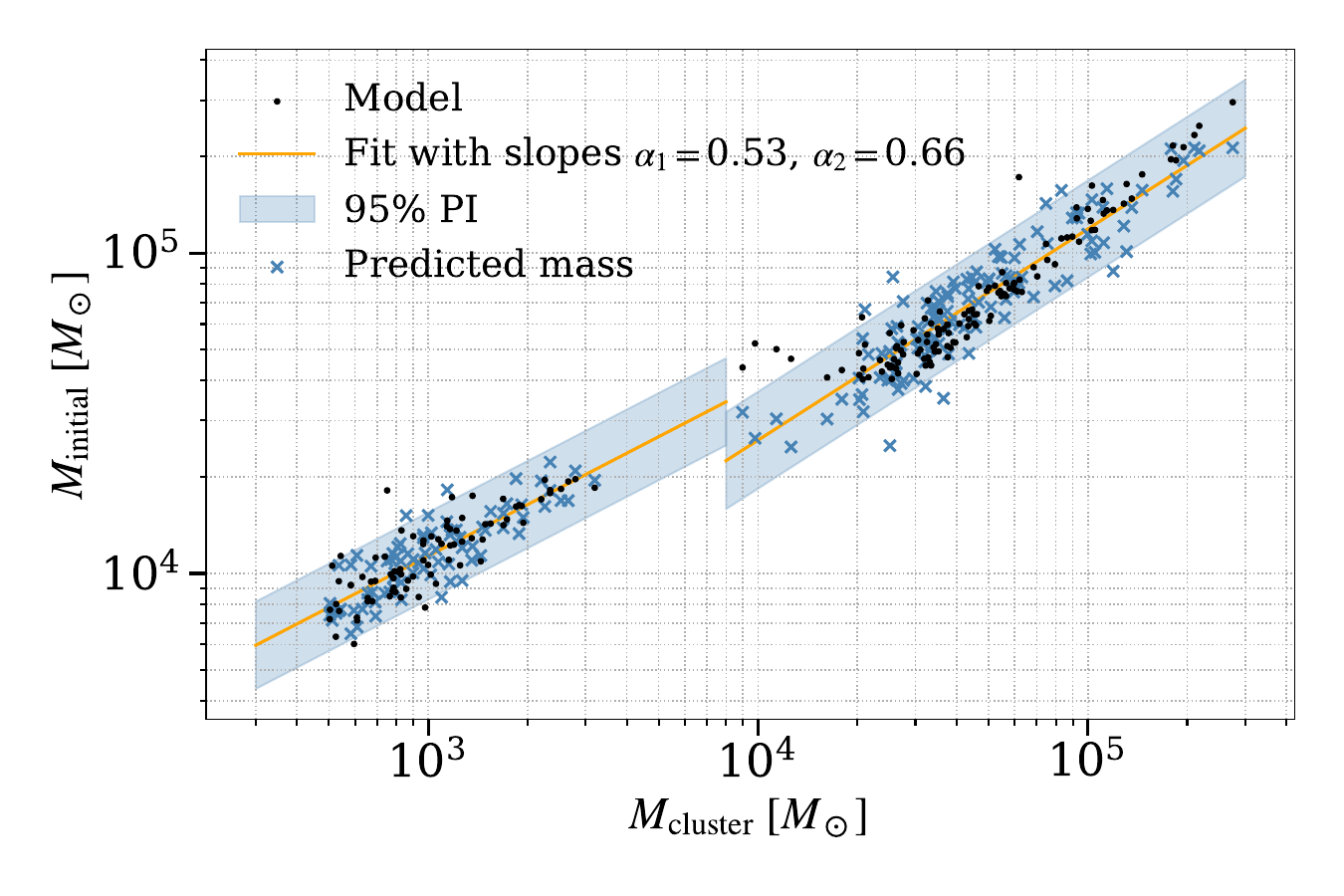}
\caption{Log-log fit to the simulated initial cluster mass and the present-day cluster mass using weighted OLS regression in two mass bins (boundary at $M_{\rm cluster}=8000\,M_{\odot}$). Similar to Figure~\ref{fig:predictm_stream}, only systems with $L>1000\,{\rm kpc\,km\,s^{-1}}$ are included for the fits. The lower-mass fit has a slope $\alpha_1=0.53$, and the higher-mass fit has a slope $\alpha_2=0.66$. The blue band shows the 95\% prediction interval, and the crosses show the possible predicted initial cluster mass based on the fit and the 95\% interval.}\label{fig:predictm_cluster}
\end{center}
\end{figure}

Similarly, to determine the relationship between remnant cluster mass and their initial mass, we perform OLS regression between ${\rm log}\,M_{\rm cluster}$ and ${\rm log}\,M_{\rm initial}$ for clusters with $L > 1000\,{\rm kpc\,km\,s^{-1}}$ (Figure~\ref{fig:predictm_cluster}). We apply two separate log-log fits based on the present-day cluster mass, yielding slopes of $\alpha_1=0.53$ for the lower-mass bin ($M_{\rm cluster}<8000\,M_{\odot}$) and $\alpha_2=0.66$ for the upper-mass bin ($>8000\,M_{\odot}$).

\begin{figure}
\begin{center}
\includegraphics[width=\columnwidth]{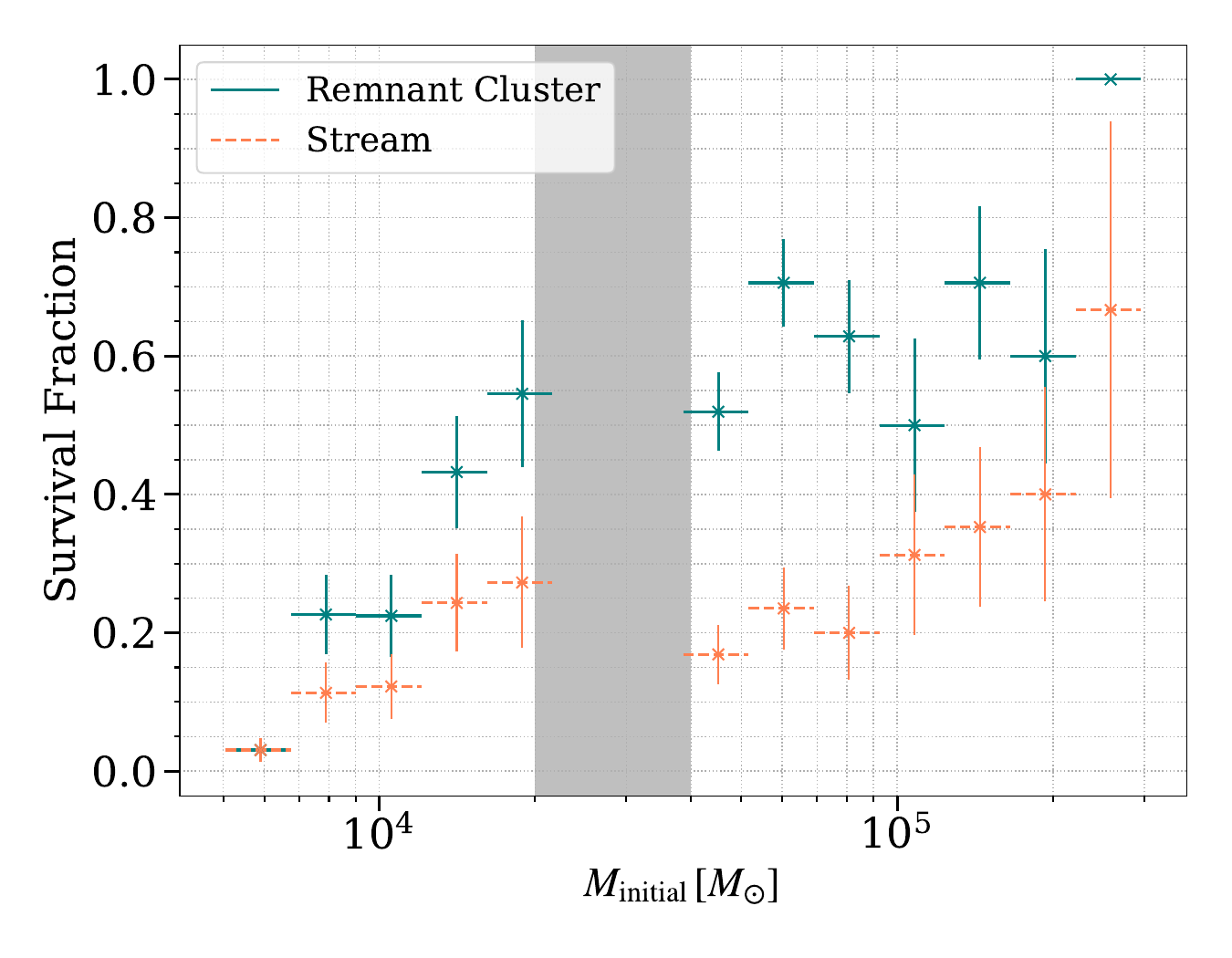}
\caption{Survival fractions $f_{\rm sv}=N_{\rm sv}/N_{\rm initial}$ of streams and remnant star clusters from cosmological simulations as a function of initial cluster mass. A system is considered to have survived if it satisfies $L > 1000\,{\rm kpc\,km\,s^{-1}}$ and meets specific criteria: survived streams must be visible at the present day, while survived clusters must maintain a remnant mass $M_{\rm cluster} > 500\,M_{\odot}$. Gray vertical band marks regions with no simulated clusters. The error bars show the Poisson error of a fraction.}\label{fig:survfrac}
\end{center}
\end{figure}

In addition to the final-to-initial mass fits, the survival fractions $f_{\rm sv}=N_{\rm sv}/N_{\rm initial}$ of remnant clusters and streams also affect the inferred ICMF. Here $N_{\rm initial}$ is the initial number of progenitor star clusters. A surviving stream is one that is visible at the present day and has angular momentum $L > 1000\,{\rm kpc\,km\,s^{-1}}$ (Section~\ref{sec:stream_sim}). A surviving cluster has $M_{\rm cluster}>500\,M_{\odot}$ and $L > 1000\,{\rm kpc\,km\,s^{-1}}$. We plot the survival fractions for different initial mass bins in Figure~\ref{fig:survfrac}. As expected, the survival fraction is larger for higher-mass star clusters. Almost all of the most massive star clusters (e.g., $\gtrsim 2\times 10^5\,M_{\odot}$) survive to the present-day and a large fraction of them produce visible streams. On the other hand, only $\lesssim 10\%$ initial star clusters with mass $\lesssim 10^4\,M_{\odot}$ produce a visible stream and retain an intact star cluster over a Hubble time.

We validate our approach by combining the fits for streams and remnant clusters with their respective survival fractions to reconstruct the ICMF originally adopted in the simulations. We apply the fits to the masses of visible streams and remnant star clusters to estimate their initial masses. For each initial mass bin (same as Figure~\ref{fig:survfrac}), we account for the dissolved population by uniformly sampling additional initial cluster masses based on the survival fraction and the number of recovered clusters. Specifically, the number of additional samples required to account for fully disrupted clusters is $N_{\rm additional} = N_{\rm sample}(1 - f_{\rm sv}) / f_{\rm sv}$. This procedure compensates for clusters that have been completely disrupted and lack observable remnants, such that the total initial number of clusters is $N_{\rm tot} = N_{\rm sample} + N_{\rm additional}$. Note that to reproduce the gap between approximately $2\times10^4\,M_{\odot}$ and $4\times10^4\,M_{\odot}$ in the simulations' ICMF, we manually resample the initial cluster masses in the adjacent mass bins on either side of the gap.\footnote{A negligible fraction of the reconstructed initial masses ($<2\%$) falls outside the range of the ICMF adopted in the simulations. For these cases, we reassign the initial mass by sampling from the nearest available mass bin in the simulations.} To mitigate the effects of small-number statistics, we perform 50 independent sampling realizations and normalize the final ICMF by the number of draws. 

\begin{figure}
\begin{center}
\includegraphics[width=\columnwidth]{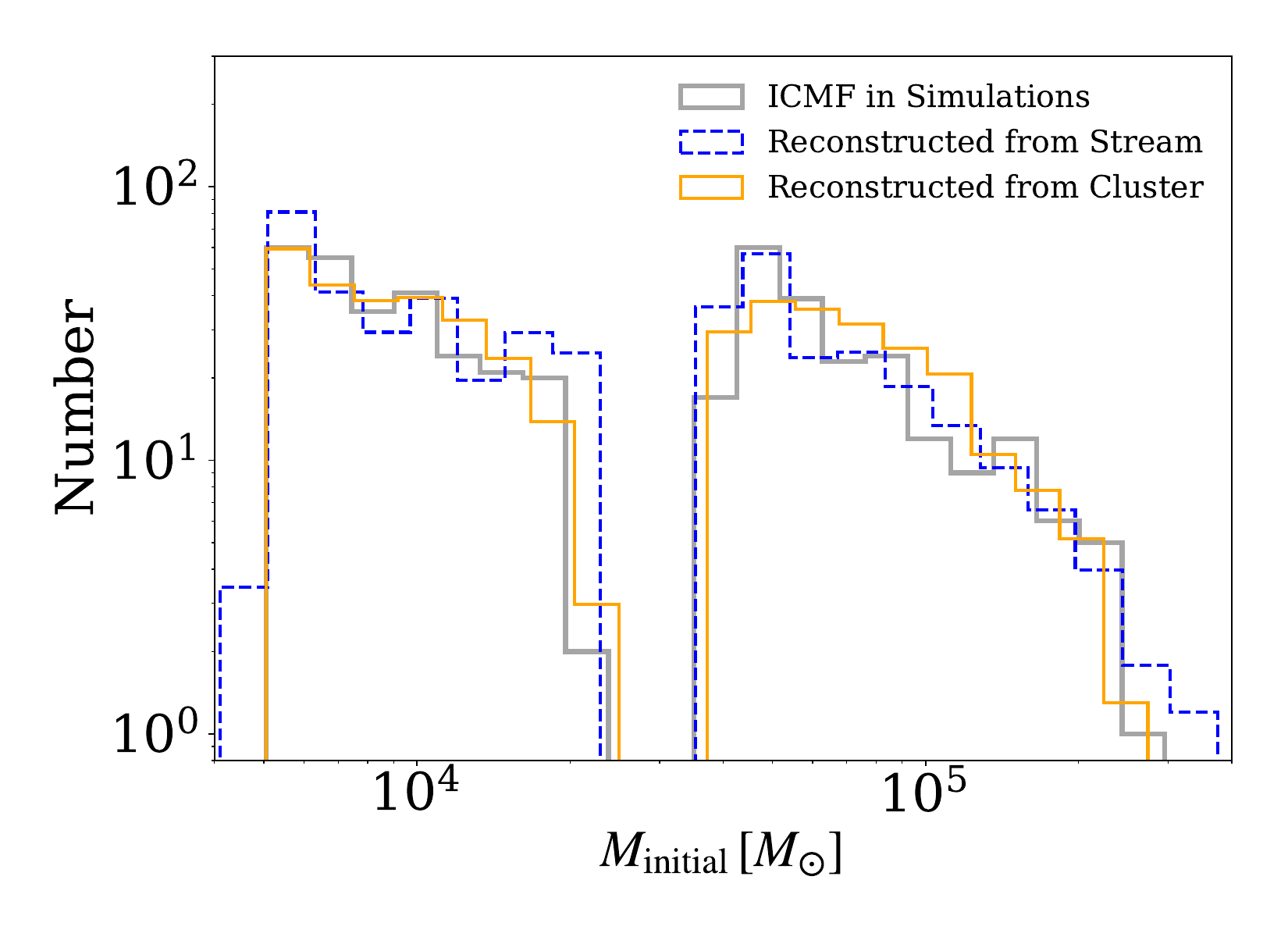}
\caption{Comparison between the adopted initial cluster mass distributions in the simulations and the reconstructed initial mass distributions inferred from the regression fits and the survival fractions.}\label{fig:predictm_imf}
\end{center}
\end{figure}

Figure~\ref{fig:predictm_imf} compares the reconstructed ICMFs from both populations to the input distribution. The KS statistics for the ICMFs recovered from simulated streams and remnant star clusters at the present day are $\approx 0.059$ and $0.06$, respectively, indicating that the reconstructed distributions are in agreement with the original ICMF.

\section{Initial Mass Function from Observed Streams and Globular Clusters}\label{sec:prediction}
\begin{figure}
    \centering
    \includegraphics[width=\columnwidth]{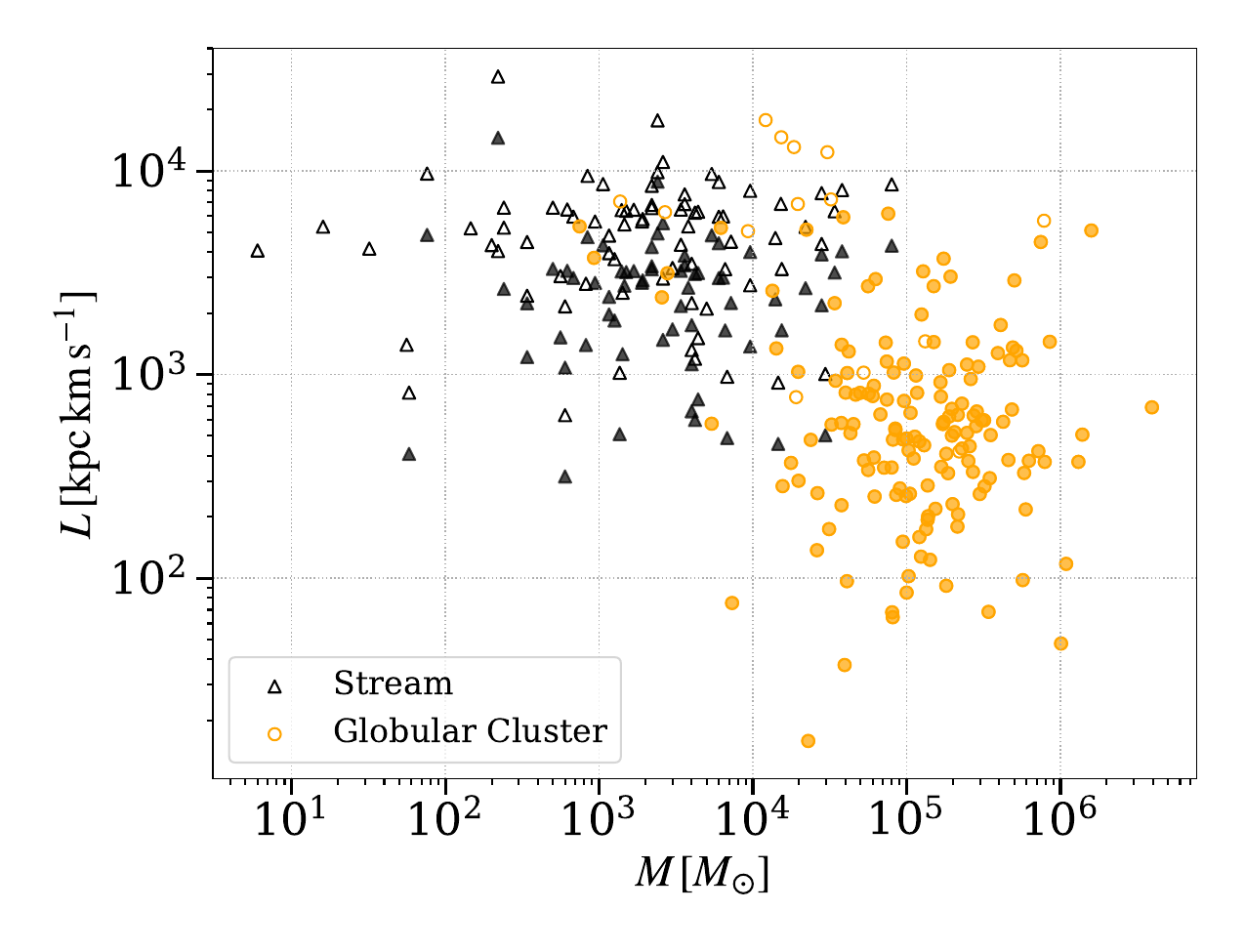}
    \caption{The angular momentum as a function of mass for stellar streams \citep[78 systems;][]{Bonaca_PW_2025} and globular clusters (165 systems; https://people.smp.uq.edu.au/HolgerBaumgardt/globular/) observed in the Milky Way. We exclude stellar streams with a dwarf galaxy progenitor. Filled triangles represent stellar streams at Galactocentric distance $D_{\rm gc} < 30~$kpc and heliocentric distance $D_{\rm hc}>3~$kpc. The observed mass of each stellar stream is corrected by a multiplicity factor of two to account for the total underlying stream mass. Similarly, filled circles denote globular clusters within 30~kpc of the Galactic center.
    }
    \label{fig:stream_m_l}
\end{figure}

In this section, we apply the fits and survival fractions in Section~\ref{sec:estimate_mini} to estimate the initial cluster mass of stellar streams and globular clusters observed in the Milky Way. Figure~\ref{fig:stream_m_l} shows the observed angular momentum versus mass of these systems, with stream data taken from \citet{Bonaca_PW_2025} and globular cluster data taken from https://people.smp.uq.edu.au/HolgerBaumgardt/globular/ \citep{Baumgardt_Vasiliev_2021, Vasiliev_Baumgardt_2021}. We exclude stellar streams with identified dwarf galaxy progenitors, resulting in 78 streams and 165 globular clusters with mass and angular momentum information in the Galaxy. Note that we account for unobserved stream material by applying a factor of two to the observed stream masses since \citet{Bonaca_PW_2025} reports mass measurements as lower limits.  In general, observed globular clusters are more massive than observed stellar streams, whereas streams typically possess larger angular momenta.

To account for observational incompleteness, we select globular clusters and streams within $D_{\rm gc} = 30$~kpc of the Galactic center for the estimates of the initial mass function. We also limit streams to those with heliocentric distance $D_{\rm hc}>3~$kpc. The observational selection effects on potential streams within 3~kpc of the Sun may be significant; for example, their massive angular span makes tracing coherent structures difficult, while the Galactic disk introduces crowding and contamination. All observed globular clusters and stellar streams used in this estimate have angular momentum $L > 1000 \, {\rm kpc \, km \, s^{-1}}$ to avoid the region of parameter space where a large fraction of systems are likely to have been disrupted during the evolution of the Galaxy. In addition, we correct for measurement uncertainties in the stream mass by applying a multiplicative factor of two (see above). As mentioned in Section~\ref{subsec:ols}, we make 50 realizations to reduce fluctuations and normalize the final ICMFs by the number of realizations.

\begin{figure}
\begin{center}
\includegraphics[width=\columnwidth]{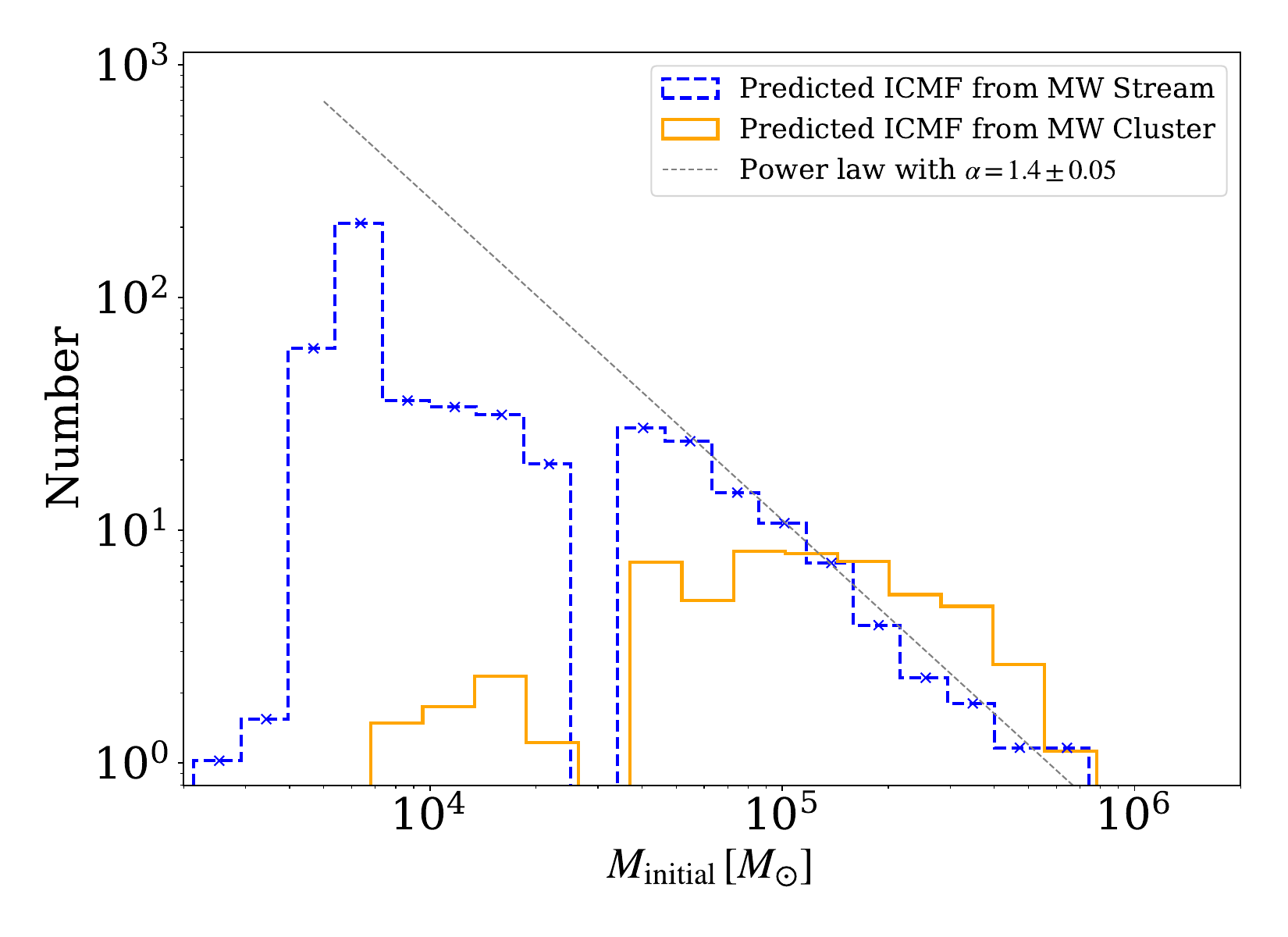}
\caption{Inferred initial mass function for a sample of Milky Way globular clusters and stellar streams. The globular clusters used for this calculation have Galactocentric distances $D_{\rm gc} < 30$ kpc. The selected streams also have $D_{\rm gc} < 30$ kpc, along with a heliocentric distance $D_{\rm hc} > 3$ kpc. All systems included in the sample require an angular momentum $L > 1000 \, {\rm kpc \, km \, s^{-1}}$. The dashed line indicates a power-law fit to the high-mass end of the ICMF ($M_{\rm initial} > 3 \times 10^4 \, M_{\odot}$) inferred from observed streams, yielding a slope of $\alpha = 1.4 \pm 0.05$.}\label{fig:predict_imf_observ}
\end{center}
\end{figure}

Figure~\ref{fig:predict_imf_observ} shows the ICMFs estimated from these observed Milky Way stellar streams and globular clusters. The higher-mass end of the predicted ICMF from observed streams follows a power-law structure, with a slope $\alpha\approx1.4$ and a standard deviation around 0.05. The lower-mass end of the stream-inferred ICMF exhibits a deficit of systems compared to the power-law fit derived from the high-mass end; this may stem from the observational incompleteness of low-mass streams. The power-law slope of the ICMF is flatter than those derived for very young star clusters ($\lesssim 1~$Gyr) in nearby galaxies, which typically exhibit a slope close to $2$ \citep[e.g.,][and references therein]{Zhang_Fall_1999,Bik+2003,McCardy_Graham_2007,PZ10,Krumholz+2019}. Most recently, JWST observations of the Antennae galaxies have identified 45 young, massive star clusters ($\lesssim 2.5~$Myr and $> 3 \times 10^4\,M_{\odot}$) with a mass function following a power-law slope of $\alpha = 1.8 \pm 0.1$ \citep{Chandar+2026}. This is consistent with the canonical slope of $\approx 2$ suggested by previous studies of more evolved populations. These results imply that early, metal-poor environments were more conducive to the formation of high-mass clusters, leading to an ICMF that is significantly flatter than those observed in younger, metal-rich cluster populations today.

In contrast, the ICMF inferred from present-day globular clusters remains inconclusive due to small-number statistics. This stems from our selection criteria, as the majority of observed globular clusters possess angular momenta below the $1000\,{\rm kpc\,km\,s^{-1}}$ threshold required for this analysis.

\section{Discussion and Uncertainties}\label{sec:uncer}
We have shown that the number and properties of observed stellar streams can be used to reconstruct the initial mass function of their progenitor clusters, providing a novel probe into star cluster formation in the early Universe. The predicted ICMF with $\alpha \approx 1.4$ in this work is flatter than the observed ICMFs in many nearby galaxies, indicating potentially different cluster formation histories in the local versus the early Universe. For example, the high-pressure environments of high-redshift galaxies may form more massive clusters than the low-pressure environments seen locally \citep[][]{Kruijssen_2026}. This evolution has broad implications for the dynamical evolution of merging binary black holes; because more massive clusters have higher escape velocities, they are better able to retain merger remnants, leading to a larger proportion of hierarchical mergers \citep[e.g.,][]{Fragione_Rasio_2023}. These systems are typically characterized by unequal mass ratios and higher black hole spins. A flatter ICMF at high redshift would thus correspond to a higher frequency of hierarchical mergers, a trend suggested by recent gravitational wave observations of a spinning subpopulation that grows more prominent at higher redshifts \citep[][]{Farah+2026}. 

The cluster dissolution mechanism in the cosmological simulations described in Section~\ref{sec:stream_sim} is designed to be consistent with detailed direct $N$-body simulations (\citealp{Carlberg18} and their Figures~4 and 5; \citealp{Carlberg24}) where stream dispersal is strongly dependent on the assumed mass assembly history of the Milky Way. We adopted simulations that have a rapid early assembly with a few major mergers in the first 5-8~Gyr, after which mass growth proceeds via subsequent halo accretion and disk star formation (modeled as prescribed buildup). In particular, a more violent buildup of the Galaxy than what is captured in our simulations would result in a larger fraction of streams being completely disrupted, meaning that the reconstructed ICMF here would underestimate the number of low-mass clusters. Future simulations exploring alternative assembly histories and other physical parameters will be essential for testing how the timing and intensity of Galactic assembly events influence stream morphology and ICMF. We note that mass loss from stellar evolution is not accounted for in the cosmological simulations. Instead, the simulations begin after massive stars have completed their evolution and all gas has been removed. This likely introduces only a factor of a few systematic uncertainties in the cluster mass.

One of the main uncertainties in the analysis arises from small-number statistics on both the simulation and observation sides. For example, there are only $\sim 10$ simulated streams with mass below $2000\,M_{\odot}$ and angular momentum above $1000\,{\rm kpc\,km\,s^{-1}}$, which naturally results in an OLS fit with a narrow prediction interval. Similarly, current stream observations are incomplete, particularly at distances beyond $\sim 20$ kpc, with significant uncertainties remaining in both the total census of stellar streams and their individual masses \citep[e.g.,][]{Bonaca_PW_2025}. In particular, the inferred slope of the ICMF depends on the measured stream mass, which introduces another main uncertainty due to observational incompleteness \citep[e.g.,][]{Jarvis+2026}, contamination of non-member stars, and assumptions about the stellar mass function. If we strictly adopt the lower-limit mass measurements without applying a multiplicity factor, the resulting ICMF slope becomes $\sim 1.6$. On the other hand, assuming an order of magnitude underestimate in stream mass measurements leads to a power-law-like slope of $\sim 0.9$. These are still flatter than the slope inferred from young star clusters as discussed in Section~\ref{sec:prediction}, confirming our conclusions. However, this also suggests that the exact slope of the ICMF depends strongly on the mass measurements of stellar streams. More complete observations from the \textit{Euclid} telescope \citep{Laureijs+2011}, the Vera C. Rubin Observatory \citep{Ivezic+2008}, and the upcoming Nancy Grace Roman Observatory \citep[][]{Spergel+2013}, will enable the detection of faint stellar streams beyond 100~kpc and stricter constraints on stream masses. These data, combined with a broader suite of cosmological simulations, will provide a more robust test for the ICMF at higher redshift.

\section{Conclusions}\label{sec:conclu}
We have presented the first analysis estimating the progenitor star cluster masses of stellar streams observed in the Milky Way. By utilizing cosmological simulations that trace globular cluster dissolution throughout the Galaxy’s assembly history, we mapped the relationship between present-day streams and their progenitors using least-square regression. Applying these regression fits and survival fractions to the observed Milky Way population, we predict an ICMF slope of $\alpha=1.4\rm0.05$ for streams with masses $\gtrsim 1000\,M_{\odot}$. While results for Galactic globular clusters and lower-mass streams remain inconclusive due to small-number statistics, the high-mass stream population suggests that the ICMF in low-metallicity, high-redshift environments may be much flatter than that of young star clusters in the local Universe. This implies that the formation efficiency of massive star clusters could be significantly higher in the early Universe. 

\begin{acknowledgments}
We thank the anonymous referee for their helpful comments and suggestions. C.S.Y. acknowledges support from the Alfred P. Sloan Foundation.
\end{acknowledgments}

\bibliography{stream_mass}{}
\bibliographystyle{aasjournalv7}

\end{document}